\documentstyle[12pt]{article}
\begin{document}
\title{A NEW APPROACH TO LOCALITY AND CAUSALITY$^+$}
\author{B.G. Sidharth$^*$\\ Centre for Applicable Mathematics \& Computer Sciences\\
B.M. Birla Science Centre, Hyderabad 500 063 (India)}
\date{}
\maketitle
\footnotetext{$^+$ Paper presented at the Vigier Symposium, Canada, 1997\\
$^*$E-mail:birlasc@hd1.vsnl.net.in}
\begin{abstract}
In the light of some recent results, it is argued that usual concepts
of causality and locality are approximations valid at scales greater
than the Compton wavelength and corresponding time scales. It follows
that the "spooky" non-locality of Quantum Mechanics is not really so
and infact is perfectly consistent with a recently discussed holistic model, which
again is corroborated by latest astrophysical and cosmological
observations. This approach
also provides a rationale for the origin of the metric and points to, what
may be called a space time quantization which may be, ultimately,
fundamental.
\end{abstract}
\section{Introduction}
In classical physics, causality has broadly three meanings\cite{r1}. 1) Causality
as predictability, what may be called Newtonian causality. 2) The fact that
no signal can have a superluminal velocity. 3) The fact that advanced effects
of fields with a finite propagation velocity are forbidden (that is the future
cannot affect the present or the past).\\
Quantum Mechanics retains the tenets of special relativity, but we speak of
microspic causality\cite{r2}: 4) The fact that observables separated by a space
like interval can be simultaneously measured, more specifically they commute.\\
However it must be borne in mind that the space time of classical physics is
not only deterministic, but it is also meaningful to speak in terms of
definite points of space time. Quantum Theory on the other hand is not only
probabilistic, but the Heisenberg Uncertainity Principle forbids the notion
of a single point space time event: Four dimensional space time exists only
as a classical approximation\cite{r3}. Inspite of this apparent contradiction,
it is still possible to give a "local" and "causal" formulation of Quantum
Theory evidenced by the fact that we deal with finite order differential
equations (cf. ref.\cite{r2}).\\
This reconciliation not withstanding the inherent contradiction between Quantum
Theory and classical physics remains and has been articulated by for example,
the EPR paradox. At the root is the issue of Quantum Mechanical acausality
and nonlocality. The acausal nature of a Quantum Mechanical measurement, according
to Einstein, violated what may be called "local realism"\cite{r4}: Individual
elements of physical reality of a system are independent of measurements
performed on any other system separated by a space like interval, that is
not in direct causal interaction. On the other hand this is opposed to a
feature of Quantum Theory, namely nonseparability\cite{r5}, according to
which two systems which interacted once cannot be assigned separate state
vectors, whatever the spatial separation. This according to Schrodinger was
"the characteristic of Quantum Mechanics".\\
In what follows we argue, in the light of some recent work, that the concepts
of locality and causality are valid only at energies and momenta greater
than those corresponding to time and space scales $\hbar/mc^2$ and $\hbar/mc$
(the Compton wavelength) but breakdown as we approach smaller space time
intervals. In the process we obtain a rationale for quantum nonseparability
and argue that the "spooky" EPR-Aspect result may not be that "spooky" after
all.
\section{Quantum Mechanical Non Locality I}
In Quantum Mechanics it is known that due to the Heiseberg Uncertainity
Principle non local effects can exist within the Compton wavelength\cite{r6}.
Indeed a recent model interprets the elementary particles as what may be
called Quantum Mechanical Black Holes, described by a Kerr-Newman type metric
with a horizon at the Compton wavelength and wherein a naked singularity
is shielded by the Zitterbewegung effects\cite{r7,r8,r9}. Within the Compton
wavelength region we have non local effects characterised by a non Hermitian
position operator. As pointed out there, physics with the conventional space
time begins at scales greater than the Compton wavelength and at time scales
greater than $\hbar/mc^2$. The result is deduced for the idealized case of
an isolated particle and could be modified when interactions are included,
though it would remain true in spirit.\\
It is within this domain that concepts of locality and causality in the usual
sense apply. Infact in Quantum Electrodynamics propagation with velocities
less than, equal to or greater than the velocity of light $c$ is allowed.
However velocities less than or equal to $c$ have overwhelmingly far greater
probability\cite{r10} at larger distances on a microscopic scale. Indeed
even in Classical Electrodynamics superluminal velocities appear within time
intervals corresponding to the Compton wavelength (cf.ref.\cite{r1}).\\
All this could be understood from a slightly different perspective. It is well
known that the Quantum Mechanical wave function which should provide as
complete a description of the system, as is possible in principle, is complex
because of the requirement of predictability by the correspondence principle\cite{r11}
The description of the wave function as
$$\psi = Re^{\imath S}$$
leads to the hydrodynamical formulation\cite{r12,r13}, from where as discussed
in detail in ref.\cite{r8}, we could get
quantized vortices with circulation velocity that of light which can be
identified with the Quantum Mechanical Black Hole referred to above.\\
That is, the requirement of predictability leads to complex Quantum Mechanical
wave functions which leads to the above model in which superluminal velocities
are relegated to a physically inaccessible region within the Compton wavelength.
Luminal and subluminal velocities are encountered in the physically accessible
region outside the Compton wavelength. This is symptomatic of the fact that
in Quantum Mechanics, unlike in the classical theory, single space time points
have no physical meaning. In the Quantum Field Theory of the Dirac equation
the wave function at two space time points does not commute, and infact
$$\{\psi (x), \psi (x')\} = 0,$$
thus apparently violating microscopic causality which is expressed by
$$[\psi (x), \psi (x')] = 0$$
for space like intervals\cite{r2}. However the commutavity is restored at
distances large compared to the Compton wavelength\cite{r14}. Moreover bilinear
forms, which correspond to physical observations, do commute. Such bilinear
forms correspond to densities with averages being taken over infinitesimal
volumes. (Infact it is precisely such an averaging over a volume corresponding
to the Compton wavelength that totally delinks negative energy components of the
Dirac wave function corresponding to superluminal non local effects from the
physical positive energy components (cf.ref.\cite{r8})).\\
Thus in all these cases we see that causal physics is restored at scales
greater than the Compton wavelength.
\section{Quantum Non Locality II}
The EPR paradox, Bell's Theorem and the Aspect experiments have been much
commented upon (cf.ref.\cite{r15}). It is the non local character of Quantum
Mechanics, which was subsequently experimentally demonstrated to which
Einstein could not reconcile himself. This non-locality is slightly
different.\\
To analyse this feature we consider the following simplified experiment: Two
identical particles which are initially together, possibly in a bound state,
get separated and move along opposite directions. We do not consider the
spin of the particles which could thus be taken to be spin less. If we measure
the momentum of one particle, say $P$, then the momentum of the other particle
$Q$, should have the same magnitude but with the opposite direction both in
Classical and Quantum Theory. The "spooky" aspect of the Quantum Mechanical
experiment is that the momentum of any particle, be it $P$ or $Q$
is determined only when an actual acausal, measurement is performed on that
particle - in other words a measurement of the momentum of $P$ should not
throw any light on the momentum of $Q$, which latter can only be determined
by a separate acausal measurement.\\
We observe that the so called paradox arises because of an application of
the law of conservation of momentum which is valid both in Classical and
Quantum Theory. Let us analyse this a little more closely.\\
This conservation law in Quantum Theory arises due to the assumption of
homogeneity of space\cite{r16}. Infact we define displacement (momentum)
operators by considering an instantaneous infinitesimal shift of origin.
The instaneous nature of this spatial shift expresses the space homogeneity
property: The displacement (or momentum) operators are independent of the
particular space point. In a physical measurement, such an instantaneous shift
corresponds to an infinite (superluminal) velocity: Infact the homogeneity
of space is a non local property. In other words the conservation of momentum
law as it stands expresses a non local property, which as in this case is
compatible with a closed system (cf.ref.\cite{r16}). Alternatively this
law is valid if the instantaneous displacement can also be considered to be
an actual displacement in real time $\delta t$. This is the case when the
Hamiltonian is not an explicit function of time $t$. This is a stationary
or steady state case and it is only under these circumstances that the
space and time displacement operators are on the same footing, that is we have
a symmetry between space and time as in special relativity\cite{r17}. It
is important to bear this in mind. Infact the symmetry between space and
time has been overstated (cf.also ref.\cite{r3}). Our perception and description
of the universe is "all space" at "one instant of time". Such a description
is clearly non-local except for the steady state case.\\
The following circumstance provides further insight into the matter. In the
light of the model discussed in\cite{r9} (cf. also\cite{r18}), wherein particles are created
fluctuationally from the background Zero Point Field, the Compton wavelength
$l$ of the pion being the typical length and the pion itself a typical particle,
we observe that the fundamental Quantum Mechanical uncertainity follows
as a consequence, rather than as an apriori consideration, in the form of
the well known equation,
$$l \sim R/\sqrt{N}$$
where $R$ is the radius of the universe and $N \sim 10^{80}$ is the
number of particles, typically pions.\\
It is worth mentioning that if there are $N$ particles in a system and $R$
is its dimension, then the typical uncertainity length $l$ is given by the
above relation (cf. also.ref.\cite{r19}). In the thermodynamic limit $N \to
\infty$ this uncertainity length $\to 0$ and we are in the classical
domain. In any case, as $N$ is large the classical concept of conservation
of momentum can be taken to hold.\\
From this point of view it appears that the very Quantum Mechanical behaviour
which leads to non locality is a natural consequence of the holistic nature
of space itself with an inbuilt non-locality (cf.ref.\cite{r9} also). This will be further elaborated in
the next section.\\
In the light of the above comments, the "spooky" or non local character is no
longer so surprising, given the non-local character embodied in homogeneity
(or conservation laws) and Quantum non-separability, which, we now argue, provides an underpinning
for space.
\section{Quantum Nonseparability and Metric}
We have noted in the introduction that Schrodinger considered the nonseparability
of two wave functions as the characteristic of Quantum Mechanics. This can be
understood in terms of the "fluctuation model" (cf.ref.\cite{r9}): Particles
are created from fluctuations of a background Zero Point Field trapped within
the Compton wavelength, a model which as we will see briefly leads to a cosmology consistent with
observation. In this model the various particles are interconnected or form
a network by the background ZPF effects taking place within time intervals
$\hbar/mc^2$ and corresponding to virtual photons of QED. Infact if two
elementary particles, typically electrons, are separated by distance $r$,
remembering that the spectral density of this field is given by\cite{r20},
(cf.also ref.\cite{r9})
$$\rho (\omega) \alpha \omega^3$$
the two particles are connected by those quanta of
the ZPF whose wave lengths are $\ge r$. So the force of (electromagnetic)
interaction is given by,
$$\mbox{Force} \quad \alpha \int^\infty_{r} \omega^3 dR,$$
where
$$\omega \alpha \frac{1}{R},$$
$R$ being a typical wavelength.\\
Finally,
$$\mbox{Force}\quad \alpha \frac{1}{R^2}$$
Thus in the idealised case of two stationary isolated particles, we have
recovered the Coulomb law. This justifies Feynman's statement that action-
at-a-distance must have a close connection with field theory. (More precisely,
as pointed out in reference\cite{r8}, the Force field is given correctly by
the Kerr-Newman metric. For a somewhat similar but simpler derivation of the
Coulomb law cf.ref.\cite{r21}).\\
It is this property of interconnectivity of the particles which indeed defines a set of particles,
which is an important starting point if we do not assume, as we should not,
background space. The point is do we consider a background space as an apriori
container of matter, or do we consider the material content of the universe
itself defining space (cf.ref.\cite{r22} for a discussion). We adopt the
latter viewpoint.\\
Starting now from the set (rather than manifold) of particles as above it is
possible to define a metric. One way of doing this is by first defining the
neighbourhood of an element as a subset of some universal set of particles,
which contains the element $a$ say, and atleast one other distinct element.\\
We now assume the following property: Given two distinct elements (or even
subsets) $a$ and $b$, there is a neighbourhood $N_{a_1}$ such that $a$
belongs to $N_{a_1}, b$ does not belong to $N_{a_1}$ and also given any
$N_{a_1}$, there exists a neighbourhood $N_{a_\frac{1}{2}}$ such that
$a \subset N_{a_\frac{1}{2}} \subset N_{a_1}$, that is there exists an infinite sequence of
neighbourhoods between $a$ and $b$. In other words we introduce topological
closeness.\\
From here, as in the derivation of Urysohn's lemma\cite{r23}, we could define
a mapping $f$ such that $f (a) = 0$ and $f(b) = 1$ and which takes on all
intermediate values. We could now define a metric, $d(a,b) = |f(a)-f(b)|.$
We could easily verify that this satisfies the properties of a metric.\\
The point is, that in the usual theory, we have apparently unconnected
particles occupying the same background space. However, it is non-local linkages
within time intervals of $\sim \hbar/mc^2$ that provide the underpinning for
space itself (cf.ref.\cite{r9}).\\
A physical picture of the above consideration is obtained if we start with
a set of $n$ particles or subconstitutents. There are $2^n$ subsets and
as $n \to \infty, 2^n \to C$ where $C$ is the cardinal number of the real
continuum. As each subset defines atleast one particle or subconstituent,
we end up with a continuum in the above process even though we start off
with a countable set.\\
Two points to be emphasized are: Firstly we had to define the set of particles,
that is physically, we defined a rule which enables us to determine whether
the particle belongs to the set or not. Secondly we arrived at a metric
starting from a larger set. This holistic aspect which we encountered in the
previous section has been commented upon
in\cite{r9} (cf.also ref.\cite{r24}), and is shown to be the reason why the pion mass
is related to the Hubble constant, an otherwise inexplicable relation which
Weinberg calls mysterious (cf.ref.\cite{r6}).\\
However two points need to be emphasized here:\\
Firstly, this "mysterious" relation,
$$m_\pi = (\frac{H\hbar^2}{Gc})^{1/3}$$
actually follows from the theory in ref.\cite{r9}.\\
Secondly, the related cosmological model\cite{r9,r25}, apart from
actually deducing the large number coincidences of cosmology, predicts an ever
expanding, accelerating universe with decreasing density as has been
observationally confirmed recently\cite{r26,r27}.
\section{Quantization of space time and Quanta}
It was pointed out in the introduction that nonlocality arises in Classical
Electrodynamics within time intervals of the order $\hbar/mc^2$. This has
lead to the concept of the chronon - a minimum unit interval $\tau_o$ of
time of the same order\cite{r28}. This some what adhoc procedure eliminates
in Classical theory the runaway solutions of Dirac's equations, but
otherwise has no strong rationale. When we neglect $\tau_o^{2}$ and higher
orders we get back the usual classical equations. For time intervals smaller
than the chronon, that is roughly less than $10^{-23}$ seconds, the motion
can be random. But for larger time intervals, special relativity holds.\\
On the other hand this concept can be extended to the domain of Quantum
Theory (cf.ref.\cite{r28} and also \cite{r29}). This leads to the fact
that wave packets have a minimum spatial spread of the order of the Compton
wavelength. Indeed even in the Classical Theory of the electron, if the
minimum space spread or radius of the approximate spherical electron,
$R \to 0$, we get saddled with the well known infinities.\\
The above two separate constraints on the minimum size of intervals in space
and time emerge in a unified way in the model of a Quantum Mechanical Black
Hole as discussed elsewhere (cf.references\cite{r8,r30}). Here as mentioned
earlier the electron for example, is the Kerr-Newman type black hole, bounded
by the Compton wavelength, but with a naked singularity shielded by the
unphysical Zitterbewegung effects (reminiscent of the random motion in the
above chronon consideration), which disappear on averaging over intervals
of the order of $\tau_0$. Further in this model as above the Compton
wavelength gives a natural boundary for a wave packet: As we approach it we
encounter unphysical negative energies corresponding to non Hermitian
operators. The correct field of the electron, including the anamolous gyro
magnetic ratio $g = 2$ emerges quite naturally. The whole point is as mentioned
earlier, we cannot uncritically carry over classical concepts of space time
to the micro domain, that is to the relatively high energy domain of Quantum
Mechanics. This has been noted by a few scholars\cite{r31,r32}. (In Quantum
Gravity too, such a granulation is recognised at the Planck scale\cite{r33}.
This has not led to fruitful results though).\\
Special Relativity and related concepts of locality and causality are phenomena
at space scales greater than the Compton wavelength and corresponding time
intervals (the chronon): There is an ultimate quantization of space and time,
a granularity, which is glossed over at our usual energy scales. The
infinities we encounter in Classical and Quantum Theory are due to our
extrapolating the usual theory in to a domain in which it is no longer valid,
viz., a domain bounded by $\tau_0$ and the Compton wavelength in which
locality and causality no longer hold.\\
It is interesting to note that, given the Compton wavelength and $\tau_0$ the
velocity of light $c$ appears as the limiting velocity our physical universe
permits. If there were no such upper bound, then for a certain observer there
would be no Quantum Mechanical Black Holes, that is no fermions or material
content in the universe.\\
Infact in the Quantum Mechanical Black Hole model, as seen earlier the particles are created
from the Zero Point Field within the Compton wavelength which is a cut
off (cf.ref.\cite{r9}) - the spectral density of the ZPF itself being
$\alpha \omega^3$, where $\omega$ is the frequency, a relation which is
compatible with Special Relativity (cf.ref.\cite{r20}). Thus we see in this
picture, a convergence of Special Relativity and Quantum Mechanics.\\
One could consider the minimum space and time intervals as being more
fundamental than the quantization of energy. Indeed in elementary Classical
Theory if the wave length has a discrete spectrum, then so does the frequency
because their product equals the velocity of light. Alternatively the
frequency is inversely proportional to the time period and hence will have
a discrete spectrum. This leads to Planck's law. The derivation is similar
to the well known theory.\\
Infact let the energy be given by
$$E = g(\nu)$$
Then, $f$ the average energy associated with each mode is given by,
$$f = \frac{\sum_\nu g(\nu) e^{-g(\nu)/kT}}{\sum_\nu e^{-g(v)/kT}}$$
Again, as in the usual theory\cite{r34}, a comparison with Wien's functional
relation, gives,
$$f = \nu F (\nu/kT),$$
whence,
$$E = g(\nu)\alpha \nu,$$
which is Planck's law.\\
Yet another way of looking at it is, as the momentum and frequency of the
classical oscillator have discrete spectra so does the energy.

\end{document}